\documentclass[graybox]{svmult}

\usepackage{mathptmx}
\usepackage{helvet}
\usepackage{courier}
\usepackage{type1cm}

\usepackage{makeidx}
\usepackage{graphicx}

\usepackage{multicol}
\usepackage[bottom]{footmisc}

\makeindex

\begin{document}

\title*{
  Two-component jet simulations: Combining analytical and numerical approaches
}
\titlerunning{Two-component jet simulations}
\author{
  Matsakos T. \and
  Massaglia S. \and
  Trussoni E. \and
  Tsinganos K. \and
  Vlahakis N. \and
  Sauty C. \and
  Mignone A.
}
\authorrunning{Matsakos et al.}
\institute{
  Matsakos T., Massaglia S., Mignone A. \at DFG, University of Turin, via P.
    Giuria 1, 10125 Torino, Italy, \email{matsakos@ph.unito.it} \and
  Trussoni E. \at INAF/Osservatorio Astronomico di Torino, via Osservatorio 20,
    10025 Pino Torinese, Italy \and
  Tsinganos K., Vlahakis N. \at IASA and Section of Astrophysics, Astronomy and
    Mechanics, Department of Physics, University of Athens, Panepistimiopolis,
    15784 Zografos, Athens, Greece \and
  Sauty C. \at Observatoire de Paris, L.U.Th., 92190 Meudon, France
}

\maketitle

\abstract*{
  Recent observations as well as theoretical studies of YSO jets suggest the
  presence of two steady components: a disk wind type outflow needed to explain
  the observed high mass loss rates and a stellar wind type outflow probably
  accounting for the observed stellar spin down.
  In this framework, we construct numerical two-component jet models by properly
  mixing an analytical disk wind solution with a complementary analytically
  derived stellar outflow.
  Their combination is controlled by both spatial and temporal parameters, in
  order to address different physical conditions and time variable features.
  We study the temporal evolution and the interaction of the two jet components
  on both small and large scales.
  The simulations reach steady state configurations close to the initial
  solutions.
  Although time variability is not found to considerably affect the dynamics,
  flow fluctuations generate condensations, whose large scale structures have a
  strong resemblance to observed YSO jet knots.
}

\abstract{
  Recent observations as well as theoretical studies of YSO jets suggest the
  presence of two steady components: a disk wind type outflow needed to explain
  the observed high mass loss rates and a stellar wind type outflow probably
  accounting for the observed stellar spin down.
  In this framework, we construct numerical two-component jet models by properly
  mixing an analytical disk wind solution with a complementary analytically
  derived stellar outflow.
  Their combination is controlled by both spatial and temporal parameters, in
  order to address different physical conditions and time variable features.
  We study the temporal evolution and the interaction of the two jet components
  on both small and large scales.
  The simulations reach steady state configurations close to the initial
  solutions.
  Although time variability is not found to considerably affect the dynamics,
  flow fluctuations generate condensations, whose large scale structures have a
  strong resemblance to observed YSO jet knots.
}

\section{Introduction}
  \label{sec:intrd}

In the last few years, a promising two-component jet scenario seems to emerge in
order to explain Young Stellar Object (YSO) outflows.
Observational data of Classical T Tauri Stars (CTTS) \cite{Edw06}, \cite{Kwa07}
indicate the presence of two genres of winds: one being ejected radially with
respect to the central object and the other being launched at a constant angle
with respect to the equatorial plane (e.g. Tzeferacos et al., this volume).
In turn, CTTS outflows may be associated with either a stellar origin, or a disk
one or with both wind components having roughly equivalent contributions.
In addition, such a scenario is supported by theoretical arguments as well (e.g.
\cite{Fer06}).
An extended disk wind is required for the explanation of the observed YSO mass
loss rates, whereas a pressure driven stellar outflow is expected to propagate
in the central region, being a strong candidate to address the protostellar spin
down \cite{Mat08}.

The goal of the present work is to study the two-component jet scenario, taking
advantage of both analytical and numerical approaches.  
In particular, we construct numerical models by setting as initial conditions a
mixture of two analytical YSO outflow solutions (each one describing a disk or a
stellar jet), ensuring the dominance of the stellar component in the inner
regions and of the disk wind in the outer.
The combination is achieved with the introduction of few normalization and
mixing parameters, along with enforced time variability of the stellar
component.
We investigate the evolutionary properties, steady states and the features of
the final configurations of the dual component jets.
Although the detailed launching mechanisms of each component are not taken into
account, such models seem capable to capture the dynamics and describe a variety
of interesting scenarios.

The employed analytically derived MHD outflows, defined as ADO (Analytical Disk
Outflow; denoted with subscript D) and ASO (Analytical Stellar Outflow; denoted
with subscript S), have been derived in the context of self-similarity
\cite{Vla98} and each one effectively describes a disk wind \cite{Vla00} or a
stellar jet \cite{Sau02}, respectively.
In Matsakos et al. \cite{Tit08}, we have addressed the topological stability, as
well as several physical and numerical properties, separately for each solution.
This article summarizes the numerical setup and reports the results of few
significant cases of the dual component jet.
A thorough study can be found in Matsakos et al. \cite{Titsu}.

\section{Numerical two-component jet models}

The two-component jet model parameters can be classified in two categories.
The first one contains those associated to the relative normalization of the
analytical solutions, i.e. the ratios of the characteristic scales of each model
(denoted with subscript *, calculated on a specific fieldline at the Alfv\'enic
surface):
\begin{equation}
  \ell_L = \frac{R_*}{r_*}\,,\quad
  \ell_V = \frac{V_{S*}}{V_{D*}}\,,\quad
  \ell_B = \frac{B_{S*}}{B_{D*}}\,,
\end{equation}
where $R_*$ is the Alfv\'enic spherical radius of the ASO model, $r_*$ is the
cylindrical radius of the Alfv\'enic surface of the ADO model (of a specific
fieldline) and the subscripts $L$, $V$ and $B$ stand for length, velocity and
magnetic field, respectively.
We assume that the protostar has a solar mass and a radius of $\sim0.01\,AU$.
Since the disk wind launching region lies in the range $0.2 - 3\,$AU, we derive
$\ell_L = 0.1$ and $\ell_V = 5.96$.
On the other hand, we define $\ell_B = 2$, which is the parameter controlling
the relative dominance of each model.

The second class of parameters concerns the mixing.
In particular, we choose the combination to depend on the magnetic flux function
$A$, which labels the fieldlines of each analytical model ($A_D$ or $A_S$).
Therefore, we define a common trial magnetic flux $A_{tr} = A_D + A_S$ and then
all physical variables $U$ are initialized with the help of the following mixing
function:
\begin{equation}
  U_{2comp} =
    \left\{1 - \exp\left[-\left(\frac{A_{tr}}{qA_{m}}\right)^d\right]\right\}U_D
    + \exp\left[-\left(\frac{A_{tr}}{qA_{m}}\right)^d\right]U_S\,,
  \label{eq:mixing}
\end{equation}
where $A_{m}$ is a constant corresponding to the matching surface rooted at
$0.16\,$AU, $q$ is a parameter that effectively moves this surface closer to the
protostar and $d$ sets the steepness of the transition from the inner ASO to the
outer ADO solution.
We choose $q = 0.2$ and $d = 2$, whereas a complete parameter study (including
$\ell_B$) can be found in \cite{Titsu}.
Essentially, Eq.~\ref{eq:mixing} provides an exponential damping of each
solution around a particular fieldline of the combined magnetic field.

Moreover, since accretion and protostellar variability are expected to introduce
fluctuations we multiply the inflow velocity with the following function:
\begin{equation}
  f_S(r, t) = 1 + \frac{1}{2}\sin\left(\frac{2\pi t}{T_{var}}\right)
    \exp\left[-\left(\frac{r}{2r_m}\right)^2\right]\,,
  \label{eq:stellar_perturbation}
\end{equation}
where $T_{var}$ is the period of the pulsation and $r_m$ is roughly the
cylindrical radius at which the matching surface intersects the lower boundary
of the computational box.
Outflow variability produces the formation of knot-like structures: the
introduction of radiation cooling (Tesileanu et al. this volume) will allow
direct comparison with observational data.

The simulations are performed with PLUTO\footnote{A versatile shock-capturing
numerical code suitable for the solution of high-Mach number flows. Publicly
available at \texttt{http://plutocode.to.astro.it}} (Mignone et al.
\cite{Mig07}).
A uniform resolution of 256 zones for every $100\,$AU is used whereas the
simulations have been carried out up to a final time of $80\,$y.
On the lower boundary we keep fixed all variables to their initial values, on
the axis we apply axisymmetric boundary conditions and at the upper and right
borders of the domain we prescribe outflow conditions.

\section{Results}

\begin{figure*}
  \resizebox{\hsize}{!}{\includegraphics{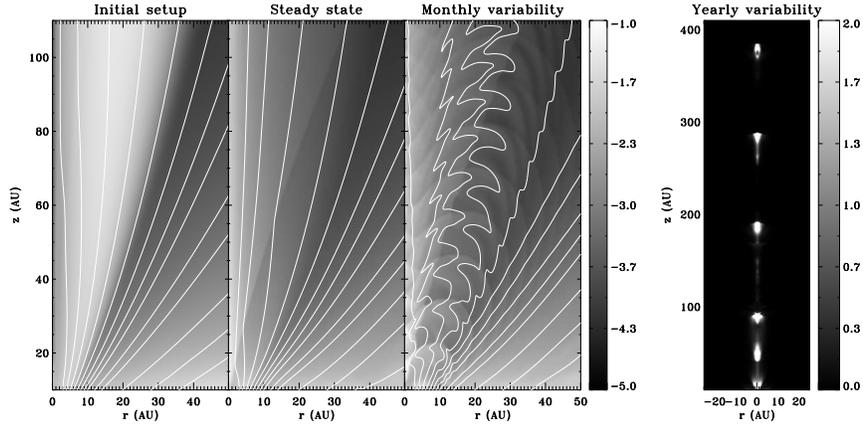}}
  \caption{
    Left panel: Logarithmic density contours (the code unit is
    $10^{-12}\,$g$\,$cm$^{-3}$) and magnetic fieldlines for the initial
    two-component jet model (left), final steady state (middle) and when a
    monthly flow variability is applied (right).
    Right panel: The quantity $10^3\rho^2\sqrt{T}$ (roughly related with
    emissivity) is plotted for the yearly variable model.
    Although $\max(10^3\rho^2\sqrt{T}) = 53.9$, the color bar uses a lower
    maximum value to enhance the displayed features.
  }
  \label{fig:figure}
\end{figure*}

In the left panel of Fig.~\ref{fig:figure}, the initial setup (left) and the
final configuration (middle) of the two-component jet are displayed.
The model shows remarkable stability and reaches a well defined steady state in
only a few years.
In particular, the disk wind solution remains almost unmodified whereas the
stellar component gets compressed around the axis.
Moreover, a shock manifests during time evolution, located roughly along the
diagonal line which crosses (10, 40) and (30, 100) (steady state plot).
This shock is found to causally disconnect the acceleration regions from the jet
propagation physics and the subsequent interaction with the outer medium.
Note that there is no such ``horizon'' present in the initial setup.
Furthermore, on the right plot of the left panel of Fig.~\ref{fig:figure}, the
same model is displayed when a monthly time variable velocity is applied on the
lower boundary (Eq.~\ref{eq:mixing}).
Evidently, despite the strong gradients seen in the density and the wiggling of
the magnetic fieldlines, the general structure is retained, proving the
stability of the two-component jet model.

The fact that the system remains very close to the initial configuration,
demonstrates that the analytical solutions provide solid foundations for
realistic two-component jet scenarios.
Consequently, specific YSO systems can be addressed more accurately by
constructing analytical outflow solutions with the desirable characteristics,
before merging them into a two-component regime.

On the right panel of Fig.~\ref{fig:figure} a quantity related to emissivity is
plotted in larger scales when a yearly variability is applied.
Near the base, the numerical solution remains close to the initial ADO and ASO
models.
However, farther away along the flow the fluctuations create knot-like
structures, which may be related with jet variability.
In fact, note that the model is associated with a condensations spacing
$\sim100\,$AU, similar to the knot spacing of HH30.

Finally, although not presented in this article, an other important parameter is
the one controlling the relative contribution of each component, $\ell_B$, with
which we can effectively and smoothly switch the model from a totally
magneto-centrifugal wind to a pressure driven jet \cite{Titsu}.

\section{Conclusions}

To sum up (taking also into account the results of \cite{Tit08} and
\cite{Titsu}), most of the technical part concerning two-component jets, e.g.
2.5D stability, steady states, parameter study, time variability etc., is now at
hand \cite{Titsu}, providing us with all the necessary ingredients to address
YSO jets.
Namely, with a) the proper analytical solutions, i.e. desirable lever arm, mass
loss rate etc., b) the correct choice of the mixing parameters and c) an
enforced time variability that effectively produces knot structures, we are now
ready to qualitatively study different and realistic scenarios, address observed
jet properties and ultimately understand the various outflow phases of specific
T Tauri stars.

\begin{acknowledgement}
The present work was supported in part by the European Community's Marie Curie
Actions - Human Resource and Mobility within the JETSET (Jet Simulations,
Experiments and Theory) network under contract MRTN-CT-2004 005592 and in part
by the HPC-EUROPA++ project (project number: 211437), with the support of the
European Community - Research Infrastructure Action of the FP7 “Coordination and
support action” Programme.
\end{acknowledgement}

\end{document}